\documentclass[iop]{emulateapj}
\usepackage{epsfig,natbib}
\usepackage{graphicx}
\received{July 3, 2014}
\accepted{September 24, 2014} 
\slugcomment{Accepted by AJ}
\newcommand{\teff}{\ensuremath{T_{\rm eff}}}

\newcommand{\rpl}{\ensuremath{R_{\rm p}}}

\newcommand{\rearth}{\ensuremath{R_{\earth}}}

\newcommand{\eke}{\emph{Kepler\ }}
\newcommand{\ek}{\emph{Kepler}}

\newcommand{\ehste}{\emph{HST\ }}
\newcommand{\ehst}{\emph{HST}}

\shortauthors{Gilliland {\it et~al.\/}}
\shorttitle{HST Imaging of Kepler Host Stars}

\begin{document}
\pagenumbering{arabic}

\title{Hubble Space Telescope High Resolution Imaging of Kepler Small and Cool Exoplanet Host Stars\altaffilmark{*}}
\author{Ronald~L.~Gilliland\altaffilmark{1}, 
Kimberly~M.~S.~Cartier\altaffilmark{1}, 
Elisabeth~R.~Adams\altaffilmark{2}, 
David~R.~Ciardi\altaffilmark{3}, 
Paul~Kalas\altaffilmark{4}, and 
Jason~T.~Wright\altaffilmark{1} 
}
\altaffiltext{1}{Department of Astronomy and Astrophysics, and Center for Exoplanets and Habitable Worlds,
The Pennsylvania State University, 525 Davey Lab, University Park, PA 16802, USA; gillil@stsci.edu}
\altaffiltext{2}{Planetary Science Institute, 1700 East Fort Lowell, Suite 106, Tucson, AZ, 85719, USA}
\altaffiltext{3}{NASA Exoplanet Science Institute/Caltech, Pasadena, CA 91125, USA}
\altaffiltext{4}{Astronomy Department, University of California, Berkeley, CA 94720, USA}

\altaffiltext{*}{Based on observations with the NASA/ESA {\em Hubble
Space Telescope}, obtained at the Space Telescope Science Institute,
which is operated by AURA, Inc., under NASA contract NAS 5-26555.}

\begin{abstract}
High resolution imaging is an important tool for follow-up study of 
exoplanet candidates found via transit detection with the {\em Kepler Mission}.
We discuss here {\em HST} imaging with the WFC3 of 23 stars that host particularly interesting
{\em Kepler} planet candidates based on their small size and cool equilibrium
temperature estimates.  Results include detections, exclusion of background
stars that could be a source of false positives for the transits, and
detection of physically-associated companions in a number of cases providing
dilution measures necessary for planet parameter refinement.
For six KOIs, we find that there is ambiguity in which star hosts the
transiting planet(s), with potentially strong implications for
planetary characteristics.
Our sample is evenly distributed in G, K, and M spectral types.
Albeit with a small sample size, we find that physically-associated binaries
are more common than expected at each spectral type, reaching a factor of 10
frequency excess at M.  We document the program detection sensitivities,
detections, and deliverables to the {\em Kepler} follow-up program archive.
\end{abstract}



\section{Introduction}

The NASA {\em Kepler Mission} has presented a catalog of over 2000 stars with
over 2700 planet-like transit signatures \citep{boru11, bata13, burk14}.
The number of planets confirmed (much more likely to be a real exoplanet
than a false positive) from the \eke sample has reached 977, with the 
vast majority of these following from the \citet{liss14} and \citet{rowe14}
validation of planets in systems showing more than one set of periodic transits.
Perhaps more impressive than these large numbers that have qualitatively 
expanded the sample of exoplanets are the large number of specific cases
for which detailed studies have allowed qualitatively new discoveries, e.g.,
circumbinary planets \citep{doyl11}, planet properties constrained by 
transit timing variations \citep{holm10}, density constraints across the 
rocky to gaseous planet domain using radial velocities \citep{marc14}, 
and significant improvements that follow from \eke asteroseismology
results for planet hosts \citep{hube13}.  

Clearly the spectacular time-series data collected from the {\em Kepler Mission}
\citep{boru10, koch10, jenk10} have been the primary basis for new advances.
Also important has been an emphasis from the mission's start to carefully
discriminate between planet candidates as found from apparent transits,
and validated or confirmed planets.  The chance for false positives in 
transit detection experiments is high, with numerous channels 
potentially contributing apparent results that are not in fact due to 
extrasolar planets \citep{brow03}.

The \eke data are used directly
to eliminate the vast majority of false positives arising from the most
common case of blended, background eclipsing binaries
through measurement of centroid motion \citep{brys13} in difference images in- and
out-of-transit.  Although very powerful, the \eke data centroiding still
allows for both false positives to slip through and for blended stars,
whether physically associated or chance superpositions, to introduce 
dilution that needs to be corrected for in arriving at accurate 
interpretations of planets.

Fundamental to the {\em Kepler Mission}
results has been a vigorous program to obtain supporting spectroscopy
and high-resolution imaging for recognizing false positives,
for confirming planets, and for refining parameters of detected exoplanets.
Spectroscopy is particularly important for detecting tight binaries
via induced radial velocity variations, and blended objects in cross-correlations
with templates.  High resolution imaging gains relevance in picking up
wider bound components and chance superpositions \citep{mort11} important to
establishing planetary status.

The primary \ehste program, data characteristics and basic image analyses
are discussed in Section 2.  Defining point spread functions, conducting
searches, avoiding spurious detections, establishing completeness limits
for the \ehste high-resolution imaging, and placing these results in
context of other similar observations are the topics in Section 3.
Section 4 presents results on using isochrone matching to assess the
probability of spatially close stars being physically associated, and
documents all companions that could be the transit host, or source of
false positives for our sample.

\section{Program, Data and Basic Image Analyses}

In early 2012, when this GO-12893 program was proposed, the {\em Kepler Mission}
follow-up programs were relying on 3-6m telescopes and adaptive optics
or speckle imaging for high resolution imaging.
Many of the \eke targets are rather faint -- with the interesting 
M-dwarf hosts of candidates often near $V$ = 16.  With transits only
a few hundred parts-per-million (ppm), deep imaging to delta
magnitudes of at least 8 was needed to securely detect all possible false positives.
This need for many Kepler Objects of Interest (KOIs) to reach the 
equivalent of $V$ = 24 severely challenged then available ground-based 
resources.  A test \citep{gill11} showed that \ehste imaging
over half an orbit could provide superior results to full-night 
efforts from the ground for the most challenging targets.  In later seasons with 
the more capable Keck-AO imaging in routine
use some aspects of the \ehste advantage have gone away, with these
having comparable resolution, and depth limits. 
\ehste retains an advantage of
allowing imaging in optical bandpasses that are standard and well-calibrated,
allowing accurate transformation to the \eke bandpass as
needed for use in establishing dilution corrections.

The \ehste imaging program was proposed to concentrate on a small subset
of the \eke planet candidates:  (1) those that were faint with shallow
transits severely challenging available ground-based imaging resources,
(2) planet candidates that had estimated parameters \rpl~ below 2.5\rearth~ 
and equilibrium temperatures below 500 K, thus emphasizing candidates 
with a chance of being rocky and in some cases within Habitable Zones
(HZ; \citet{kast93}).
The GO-12893 SNAPshot program contained 158 targets, with the bulk being
G-type stars mirroring the overall \eke target list.  In the end, and 
contrary to advertised expectations, the executed SNAPshot visits
often corresponded to our faintest stars and longest visits.
Thus the M dwarf targets were executed at a higher
relative fraction.  \ehste observations were obtained for
only 22 of the proposed sample of 158, however those executed were among both
the most interesting and most challenging for ground-based study.
For example, our sample of 22 KOIs included 6 of 11 highlighted in the
\citet{dres13} study of M dwarf KOIs with candidates in or near the
HZ, and all 3 (KOIs 854, 1422, and 2626) identified as most favorably
placed with respect to the HZ.
The observations listed in Table 1 span 2012 October 27 to 2014 January 7;
any further visits that occur after 2014 May 1 will be processed,
with deliveries made as discussed in Section 3.3, but are not
included in this paper.

\begin{table*}
\vspace{-0.2in}
\begin{center}
\tablenum{1}
\caption{KOIs with \ehste High Resolution Imaging
\label{tab:koilist}}
\begin{tabular*}{0.7\textwidth}{rrrccrccccr}
\tableline\tableline
KIC & KOI & Kepler & N$_{p}$ & Kp & Depth & $\delta$-mag & Def$_{\rm 555}$ & Def$_{\rm 775}$ & $\teff$ & Dist \\ \hline 
2853029 & 3259 & --- & 1 & 15.68 & 872 & 7.53 & --- & 0.15 & 5494 & 1325 \\ 
4139816 & 812 & 235 & 4 & 15.95 & 554 & 8.03 & 0.17 & 0.14 & 4023 & 495 \\ 
4813563 & 1959 & --- & 1 & 14.25 & 859 & 7.55 & --- & --- & 4915 & 480 \\ 
5358241 & 829 & 53 & 3 & 15.39 & 422 & 8.32 & --- & 0.03 & 6266 & 1765 \\ 
5942949 & 2525 & --- & 1 & 15.70 & 810 & 7.61 & --- & 0.10 & 4595 & 715 \\ 
6026438 & 2045 & 354 & 3 & 15.55 & 356 & 8.51 & --- & --- & 4621 & 710 \\ 
6149553 & 1686 & FP & 1 & 15.89 & 631 & 7.89 & 0.32 & --- & 3510 & 235 \\ 
6263593 & 3049 & --- & 1 & 15.04 & 540 & 8.05 & --- & --- & 4582 & 690 \\ 
6435936 & 854 & --- & 1 & 15.85 & 1692 & 6.81 & 0.26 & --- & 3580 & 265 \\ 
7455287 & 886 & 54 & 3 & 15.85 & 448 & 8.26 & 0.22 & --- & 3654 & 290 \\ 
8150320 & 904 & 55 & 5 & 15.79 & 436 & 8.29 & 0.00 & 0.11 & 4404 & 645 \\ 
8890150 & 2650 & 395 & 2 & 15.99 & 357 & 8.50 & 0.23 & 0.10 & 3825 & 390 \\ 
8973129 & 2286 & --- & 1 & 15.06 & 504 & 8.13 & --- & --- & 5452 & 975 \\ 
9838468 & 2943 & --- & 1 & 13.85 & 213 & 9.06 & --- & --- & 6036 & 820 \\ 
10004738 & 1598 & 310 & 3 & 14.28 & 168 & 9.32 & --- & --- & 5657 & 780 \\ 
10118816 & 1085 & --- & 1 & 15.23 & 307 & 8.67 & --- & --- & 3820 & 275 \\ 
10600955 & 2227 & --- & 1 & 14.87 & 523 & 8.09 & --- & --- & 5848 & 1090\\ 
11306996 & 3256 & FP & 1 & 14.81 & 333 & 8.58 & --- & --- & 4021 & 290 \\ 
11497958 & 1422 & 296 & 5 & 15.92 & 787 & 7.65 & 0.29 & --- & 3690 & 340 \\ 
11768142 & 2626 & --- & 1 & 15.93 & 818 & 7.60 & 0.32 & --- & 3620 & 400 \\ 
12256520 & 2264 & --- & 1 & 14.48 & 334 & 8.58 & --- & --- & 5556 & 800 \\ 
12470844 & 790 & 233 & 2 & 15.34 & 962 & 7.43 & --- & --- & 5208 & 970 \\ 
12557548 & 3794 & --- & 1 & 15.69 & 3186 & 6.13 & --- & --- & 4242 & 550 \\ \hline 
\end{tabular*}
\tablecomments{KIC, and corresponding KOI numbers are given for all targets,
with the \eke planet number given for validated cases.  The number of multiple
candidates or planets (N$_{p}$), Kp, and the shallowest transit depth (Depth in ppm) over multiple systems
are taken from CFOP.  The $\delta$-mag shows how much fainter a false 
positive source could be assuming 90\% deep eclipses, and hence the relative depth to
which high resolution imaging is needed. Def$_{\rm 555}$ and Def$_{\rm 775}$
are the net shortfalls in magnitudes
in reaching standard depth due to slightly truncated exposure times.
$\teff$ is from the KIC, and Dist (in pc) is calculated
using the \ehste photometry as discussed in the text.}
\end{center}
\end{table*}

An advantage of our \ehste imaging is a very uniform data set.
All observations were acquired with a standard set of five dithered
exposures in each of the WFC3 filters F555W and F775W, chosen to sample the 
\eke bandpass optimally.  Four exposures used the {\tt DITHER-BOX}
pattern with exposure times chosen to reach 90\% of saturation for
cases with the image centered on a pixel.  A fifth exposure at six
times the length of those in the pattern, thus saturating the 
core, was added to bring up signal-to-noise (S/N) in the wings and
allow reaching delta magnitudes near 10.  For about one-half of the faintest
targets the long exposure was limited to 300 seconds in one
of the two filters (three targets were limited in both filters), thus yielding
slightly shallower depths as listed in Table 1.
The lost depth due to truncated exposure times is calculated
separately as Def$_{\rm 555}$ and
Def$_{\rm 775}$ for the two filters by using the WFC3 Exposure Time
Calculator provided by ST ScI.  This takes into account the five 
exposures in each bandpass and realistic background levels.
The deficits are defined as the difference in magnitudes required
to reach signal-to-noise of 5 at the actual exposure time compared
to the exposure time that would have been used for the full, 
standard depth.  The deficit applies to 2\arcsec radius and beyond.
At closer offsets from the target star deficits are likely smaller
since other factors enter to limit relative depth, but for discussion
we conservatively assume these apply everywhere.
Also included in Table~\ref{tab:koilist} is KIC-12557548 added from GO-12987 (PI Rappaport,
Croll et al. 2014)
for which nearly identical exposures were available.
Image FWHM values (Gaussian fit) are 0{\farcs}077 $\pm$ 0{\farcs}004
for F555W and 0{\farcs}079 $\pm$ 0{\farcs}003 for F775W with no outliers
to statistically unexpected deviations.
WFC3/UVIS consists of two backside illuminated 2048$\times$4096 CCDs
with a scale of 0{\farcs}04 per pixel, and a full well depth of about
72000 electrons.  For a centered stellar image in our bandpasses
$\sim$15\% to 18\% of the light falls on the central pixel in
F555W and F775W respectively.  Further details of WFC3 may be 
found in \citet{dres14}.

\begin{figure}
\begin{center}
\includegraphics[width=80mm]{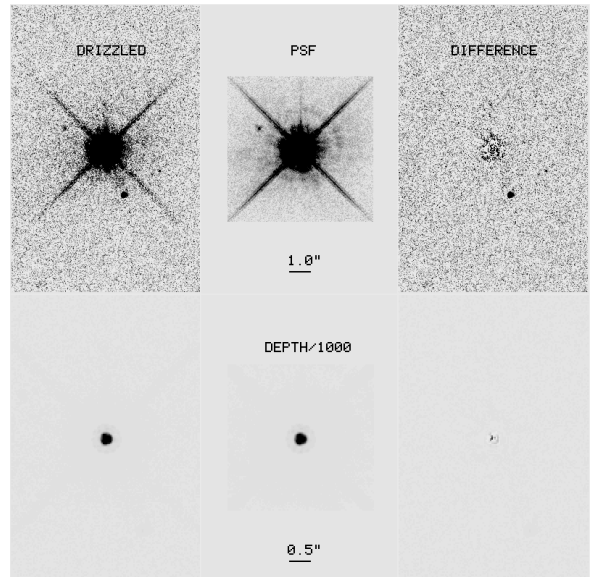}
\end{center}
\caption{KIC-10004738 related images for illustration purposes.  The left panels
show deep (upper), and shallower by $\times$1000 displays of the image drizzled
at 0{\farcs}03333 for F775W.  The middle panels show the point spread function,
as discussed in the text shown to the same stretch.  The right panels show the 
difference in sense of drizzled image minus the PSF.  The apparent
star offset by $\sim$2\arcsec~ near 10 o'clock is a filter artifact also 
represented in the PSF that disappears in the difference image, while a
true source near  $\sim$2\arcsec~ offset at 5 o'clock becomes the most obvious
feature in the difference image.  Diffraction spikes and filter artifacts are
removed to very high fidelity.  The lower panels are meant to illustrate the
inner part of the PSF, the upper panels the extended domain to $\sim$5\arcsec.}
\label{fig:diff}
\end{figure}

Figure~\ref{fig:diff} illustrates the very similar data for each target, 
KIC-10004738 in particular.

Discussion of data processing through drizzle-combinations of all available
exposures to a scale of 0\farcs03333 is presented in \citet{cartier14}.
Critical to this study the final drizzled image products were all 
shifted to be nearly (to within a few 0.01 of a WFC3 UVIS pixel)
centered on a pixel, thus allowing all images to be ``stacked" as
discussed in Section 3.1 for an empirical point spread function (PSF) definition.

The WFC3 UVIS images were obtained using the 1024$\times$1024 subarray
nearest the readout amplifier of UVIS2, thus minimizing the effects
of charge transfer efficiency.  To further minimize limitations from
the loss of charge transfer efficiency all exposures used a post-flash
(brief illumination using an LED internal to WFC3 immediately after 
the exposure, and before CCD readout)
to generate a physical background of about 12 electrons per pixel 
including allowance for the expected small sky background.

The UVIS subarray spans just over 40\arcsec$\times$40\arcsec.  A few
pixels near the boundary were trimmed to retain an area of exactly
1600 square arcseconds searched for companions to simplify later statistics.

\section{Companion Detections and Completeness}

The primary product to be obtained from these \ehste observations is
a list of all stars within the field of view, pushing as close to the
bright target as allowed by the data and PSF, and providing accurate
photometry and positions.  Associated with this will be extensive 
simulations and analyses to quantify limits to false detections, 
detection completeness as a function of depth in delta magnitudes
relative to the target star, and offset
(angular separation from the primary) distance from the target.

\subsection{Empirically Defined PSF and Subtraction}

\ehste imaging is characterized by great stability 
compared to ground-based imaging, whether direct or with adaptive optics.
However, over the course of \ehste orbits the telescope is thermally 
stressed and the focus varies by an amount that does influence the 
cores of PSFs in a significant way.  Also, through broadband filters
the PSFs will vary slightly as a function of the underlying spectral
energy distribution.  A goal of these \ehste observations is to make
use of the general imaging stability, account for deviations from this,
and thereby enable deep relative imaging near the core of the bright KOIs
by subtracting a model PSF from each individual image.

We define a model PSF for each individual image by ``stacking" all
available images, except for those in which close-in companions are
found, and then for each pixel relative to the target centers 
developing the PSF model as a function of target color, telescope
focus, and minor deviations with respect to sub-pixel phase of the
target centering.  Photometry in the standard WFC3 Vegamag system
(see Section 3.3)
was developed for each target as discussed in \citet{cartier14}, we
then adopt the F555W - F775W magnitudes as the color.
For telescope focus we utilized STScI's Observatory Support 
portal\footnote{\url{http://www.stsci.edu/hst/observatory/focus}}
that provides an estimated secondary-primary mirror despace at the 
time of observations using a modeling of observatory support structure temperature
changes.  
For residual offsets from perfect centering we used an initial 
guess based on first moments over the central 3$\times$3 pixels.
We used 20 of the 23 targets for the PSF definition, ignoring
KIC-6263593, KIC-11497958, and KIC-11768142 for which the binary
or triple nature with small delta magnitudes and 
offsets $<$0{\farcs}5 made these useless for determining a PSF.
For five of the other targets we pre-subtracted a total of six
much fainter companions that if left would have perturbed the 
resulting PSF model.  A sample of 20 images in each filter to 
define the PSF as a function of color, focus and residual centering
is smaller than desired, but clearly works well for eliminating 
to near perfection diffraction spikes and any filter artifacts.

The model PSFs in F555W and F775W are developed over an area of 
201$\times$201 of the 0\farcs03333 pixels, thus including an 
area of at least 3{\farcs}33 radius extending to 4{\farcs}7 on 
the diagonals to include the full domain of diffraction spike influence.
Although other options were experimented with, we adopted a PSF 
model that was a linear function of color, focus, and the always
small x and y offsets from perfect image-to-image alignment.

After an initial fit at each pixel to determine a zero point
plus the four coefficients of linear terms, the solution was
iterated for each image by fitting the model PSF to each individual
target providing updated linear coefficients for focus and offset
positions, with the known color being held fixed.
With revised focus and offset coefficients for each target in hand,
the model PSF at each pixel was again derived.  This was iterated
four times at which point convergence was reached.

The result of this approach can be assessed from inspection of 
Figure~\ref{fig:diff} showing one of the input images, the 
model PSF evaluated at the color, focus and offset applicable 
to this target, and the result of subtracting the scaled PSF
to create a difference image.  The PSF resulting from stacking
20 individual images shows a much lower noise level than the
single target drizzled image -- as should be the case.
Also as should be the case, easily recognized features such as
the diffraction spikes,
and a filter artifact are sufficiently
well subtracted to leave no discernible trace with careful 
inspection of difference images, while nearby faint stars are preserved.
However, within the inner $\sim$0{\farcs}2 the PSF subtraction does
not leave residuals that are near the Poisson limit, although 
detection sensitivity is gained even at these small offsets from
using the PSF model subtraction.  This lack of reaching a fundamental
limit at small offsets could follow from a number of things.
Our model for the PSF might be inadequate, e.g. perhaps other
terms than the assumed linear dependence of color, focus,
and small offsets should apply.
We did try quadratic dependence on focus with no improvement.
Perhaps the sample size of 20 targets spanning
a significant range of color (mid G through mid M stars), focus
(deviations to about $\pm$ 5 microns predicted), and residual centering imperfections
to 0.05 pixels was insufficient for a solid solution; we had
argued that 50 targets observed would be a good number for supporting
robust PSF subtractions.  In the end, though, the limiting factor
seems likely to be the extent to which each individual exposure's
position relative to the other four to be drizzled together 
could be determined.  Given the small sub-array, there are 
generally only a small handful of other stars in the field of 
view at magnitudes remotely comparable to the target from which
to establish relative positions.  Minor errors at the order of 0.01 pixels
in determining relative offsets of the under-sampled component 
dithered exposures would suffice to introduce errors to the PSF
of individual combined drizzle products that would not be 
amenable to correction through our stacking process.
Our PSF subtractions are in fact quite good, and there 
is a ready explanation for subpar performance in the core.
Given that, coupled with small sample size, we have not exhaustively 
pursued improvements in this area.  Rather, we focus on properly
quantifying the detection limits that follow from the data 
subtractions as described.

\subsection{DAOFIND Searches, Avoiding False Detections}

We use the {\tt DAOFIND} task within IRAF based on \citet{stet87}
to find stellar sources within our difference images from which
the primary target has been suppressed.
We have adopted a four-fold approach to avoiding false detections:
(1) A number of images equal to our target set (23) have been 
simulated using simple Gaussian noise at the level of 9 electrons
characterizing our real data.  We then run {\tt DAOFIND} on these
images adopting increasingly conservative cuts on the adjustable
parameter {\tt threshold} for the feature detection in sigma
until no false detections result over the full sample.
A value of {\tt threshold} = 5.0 resulted in no false detections
for this null limit test, but is recognized as perhaps not 
sufficiently conservative for the real data.
(2) Our goal is to retain most stars to a delta magnitude beyond
9.0, but to reject background galaxies to the extent possible.
We therefore generated a grid of 2500 delta-mag 9 stars at 
random sub-pixel centering phase per each of the 23 pure Gaussian
noise simulated, 1024$\times$1024 images and then plotted the distribution
of {\tt DAOFIND} output parameters on {\tt sharpness, sround}, and
{\tt ground} and adopted bounds on these as tight as possible while
eliminating no more than $\sim$1\% of these targets near the 
expected limiting depth.  We adopted a range of [0.4,0.85] for 
{\tt sharpness}, [-0.95, 0.95] for {\tt sround}, and [-0.5, 0.6]
for {\tt ground}.
(3) We performed extensive inspections of results on the actual images after
using the above parameters.  The only change following from 
this was to adopt a more conservative {\tt threshold} of 6, rather
than 5 $\sigma$, on the underlying detection significance to 
minimize detections which upon inspection seemed to not always
be confidently stellar.
(4) Near bright stars we boosted the required magnitude returned 
from {\tt DAOFIND} based on a 5th order polynomial fit to the 
mean noise in annuli spanning 0{\farcs}1 to 2{\farcs}0 away 
from target stars on average.  Without this step a large number
of spurious detections occur at small offsets in the 
difference images simply from increased Poisson noise levels, 
but also from increasingly imperfect PSF subtractions at small
offset radius.

This approach results in high completeness in detecting
any stellar companions that could result in a false 
positive for the shallow transits in question.  That is,
outside of $\sim$1{\farcs}4 where noise from the stellar
wings provides little lost sensitivity 22 of 23 cases in 
Table 1 are imaged deeply enough to exclude EBs with
90\% depth.  Closer to the stellar core at a working 
radius of $\sim$0{\farcs}4 the detection limit (adjusted
for deficits due to shortened exposures) is at about the
median of depths needed to fully probe for the worst 
case 90\% deep eclipsing binaries.
At the same time
we have been rather liberal in the retention cuts in the sense
that we likely are not pushing as near the limiting depth as 
might be possible.  In this approach we are placing a greater 
emphasis on avoiding false detections than reaching limiting 
depth, and will rely upon quantified completeness simulations
to characterize the results.  Furthermore, we will individually
discuss in Section 4 any close-in detections that are not 
unambiguously real.

In setting up the difference images upon which {\tt DAOFIND} 
is run to detect stars we subtract not only the primary target,
but also perform PSF fits to and subtract all other stars 
within a delta-magnitude of three of the primary.  This was 
found to be necessary to avoid spurious detections, e.g. along
diffraction spikes on any stars within the images.
For the three cases in which tight binaries or triples were 
detected we performed simultaneous fits of the PSF to these
to create the difference image as discussed in \citet{cartier14}.

\subsection{Photometric Results -- Deliverables to CFOP}

We use {\tt DAOFIND} only to locate the approximate position 
of stars to be analyzed.  Photometry and astrometry are then
developed using least squares fits of the PSF derived for each
image based on the discussion in Section 3.1.  
Photometry of individual stars makes use of bi-cubic polynomial
interpolation shifts of the PSFs
as solved for in the least squares fitting, with the count level
then following directly from the scale.  Formal errors on the
count levels and positions are adopted from the least-squares
fitting \citep{bevi69}.
Transformation to the VEGAmag system is based on the STScI 
published zero points\footnote{\url{http://www.stsci.edu/hst/wfc3/phot\_zp\_lbn}}
based on encircled energy in 2{\farcs}0 radii relative to an infinite aperture
along with published zeropoints.
The vega magnitude of a star with flux $F$ is -2.5 log($F/F_{vega}$) where
$F_{vega}$ is the calibrated spectrum \citep{bohl04} of Vega.
We verified that our PSF fits accurately reproduced results for 
2\arcsec~ radius aperture photometry on isolated bright stars.

\begin{figure}
\begin{center}
\includegraphics[width=80mm]{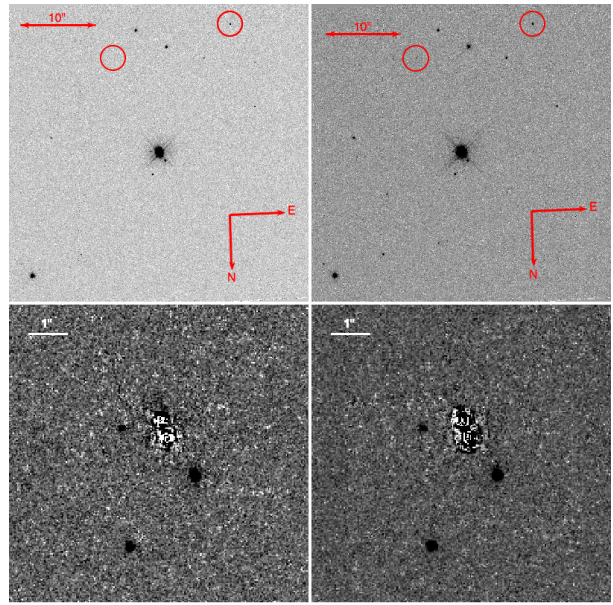}
\end{center}
\caption{
KIC-6263593 (KOI-3049) images showing the full 40$\times$40\arcsec~ field 
of view in the upper panels.  The lower panels are difference images
of an inner region after subtracting the close-in bright pair.
See Section 3.3 for further discussion.}
\label{fig:img3049}
\end{figure}

Table~\ref{tab:cfoptab} shows an example for KOI-3049, KIC-6263593
of a delivered detections table.
The photometry described in this paper is performed on the target
star and provided in the tables.  The improved target star magnitudes
take into account the contributions of the newly discovered
field and companion stars.
Our goal for this paper is not
to show all of the results for each target, but rather to document
the general approaches and highlight any interesting results.
Figure~\ref{fig:img3049} shows the drizzled images from which the
detections were developed.  Both the {\tt fits} images and these
source tables (extension of {\tt .src}) have been delivered to 
the primary archive for \eke results, CFOP\footnote{\url{https://cfop.ipac.caltech.edu}}.
For this example the delivered files are named:
{\tt 3049Ic-rg20130214v.fits, 3049Ic-rg20130214i.fits, {\rm and} \\ 
3049Ic-rg20130214HST.src};
in addition a 3$\times$3\arcsec~ clipout is available as {\tt 3049Ic-rg20130214i.jpg}.
The {\tt `v'} and {\tt `i'} in above names are shorthand for F555W
and F775W respectively.

\begin{table*}
\begin{center}
\tablenum{2}
\caption{Example Source Detection Table for CFOP Delivery
\label{tab:cfoptab}}
\begin{tabular*}{\textwidth}{ccrrccccccrrc}
\tableline\tableline
RA & Dec & Dist & PA & F775W & F775W\_e & Color & Color\_e & Kp & Kp\_e & d\_Kp & KIC & KIC\_Kp \\ \hline 
282.9975209 & 41.6785572 & 0.000 & 0.000 & 14.806 & 0.020 & 1.198 & 0.020 & 15.537 & 0.020 & 0.000 & 6263593 & 15.037 \\  
282.9974699 & 41.6784340 & 0.478 & 196.900 & 15.284 & 0.020 & 1.362 & 0.020 & 16.080 & 0.020 & 0.543 \\ 
282.9974496 & 41.6783548 & 0.757 & 194.804 & 20.592 & 0.061 & 1.280 & 0.111 & 21.355 & 0.076 & 5.818 \\ 
282.9970876 & 41.6784729 & 1.207 & 255.271 & 22.400 & 0.045 & 0.916 & 0.062 & 23.017 & 0.045 & 7.480 \\ 
282.9977796 & 41.6788370 & 1.221 & 34.638 & 19.712 & 0.015 & 1.242 & 0.022 & 20.460 & 0.018 & 4.923 \\  
282.9971164 & 41.6793261 & 2.972 & 338.496 & 20.773 & 0.019 & 1.126 & 0.027 & 21.474 & 0.022 & 5.937 \\ 
282.9936923 & 41.6799497 & 11.450 & 295.946 & 22.764 & 0.054 & 2.436 & 0.152 & 23.992 & 0.100 & 8.455 \\ 
282.9961378 & 41.6818506 & 12.423 & 342.573 & 22.942 & 0.060 & 3.001 & 0.293 & 24.397 & 0.188 & 8.860 \\ 
282.9999390 & 41.6750932 & 14.065 & 152.477 & 20.422 & 0.017 & 2.654 & 0.032 & 21.738 & 0.024 & 6.201 \\ 
282.9954610 & 41.6749528 & 14.112 & 203.117 & 22.913 & 0.046 & 3.500 & 0.721 & 24.568 & 0.458 & 9.031 \\ 
283.0023335 & 41.6769730 & 14.141 & 113.799 & 21.039 & 0.021 & 2.055 & 0.034 & 22.114 & 0.025 & 6.576 \\ 
282.9980917 & 41.6746183 & 14.266 & 173.831 & 18.003 & 0.013 & 2.083 & 0.019 & 19.090 & 0.017 & 3.552 & 6263597 & 19.349 \\ 
282.9921937 & 41.6778002 & 14.583 & 259.217 & 20.707 & 0.035 & 2.763 & 0.068 & 22.067 & 0.047 & 6.530 \\ 
282.9921506 & 41.6777921 & 14.702 & 259.189 & 20.875 & 0.042 & 2.694 & 0.080 & 22.207 & 0.055 & 6.670 \\ 
282.9974078 & 41.6831811 & 16.646 & 358.948 & 23.466 & 0.072 & 1.841 & 0.160 & 24.455 & 0.107 & 8.913 \\ 
282.9966051 & 41.6739474 & 16.780 & 188.445 & 19.552 & 0.015 & 0.619 & 0.020 & 20.050 & 0.018 & 4.513 & 6263592 & 19.659 \\ 
282.9933695 & 41.6821712 & 17.141 & 319.361 & 20.528 & 0.018 & 2.625 & 0.033 & 21.832 & 0.025 & 6.294 \\  
283.0013226 & 41.6738763 & 19.711 & 148.768 & 20.743 & 0.018 & 1.238 & 0.026 & 21.490 & 0.021 & 5.952 \\ 
282.9901758 & 41.6794287 & 19.998 & 279.017 & 23.558 & 0.072 & 2.672 & 0.330 & 24.881 & 0.212 & 9.343 \\ 
282.9909203 & 41.6828850 & 23.615 & 311.271 & 17.305 & 0.012 & 0.572 & 0.017 & 17.784 & 0.015 & 2.247 & 6263577 & 17.742 \\ 
282.9904070 & 41.6824301 & 23.669 & 306.081 & 21.840 & 0.029 & 2.158 & 0.056 & 22.956 & 0.039 & 7.429 \\ \hline 
\end{tabular*}
\tablecomments{
RA and Dec = position of \ehste source (J2000, decimal degrees) \\ 
Dist and PA = distance (arcsec) and position angle (deg. E of N) of \ehste source from KOI \\ 
F775W and F775W\_e = \ehste (Vega) magnitude and error of \ehste source \\ 
Color and Color\_e = F555W - F775W \ehst-based color and error \\ 
Kp and Kp\_e = derived \eke magnitude from F555W, F775W and error \\ 
d\_Kp = number of \eke magnitudes fainter (or brighter) than KOI in \ehst-based Kp \\ 
KIC = KepID likely corresponding to \ehste source \\ 
KIC\_Kp = KIC value of Kp for likely match \\ 
}
\end{center}
\end{table*}

\subsection{Simulations to Define Completeness}

For defining completeness we have chosen to insert an equal
number of simulated targets on each of 22 (of 46 total) images, half in the F555W
filter, and half in F775W.  We use all images for this simulation
except:  (1) The three cases of obvious binary and
triple stars; (2) the KIC-12557548 images from GO-12897,
since these went slightly deeper than the GO-12893 set;
(3) the $\sim$1/4 of images from Table 1 for which one or both filters 
were not taken to the full nominal depth.
This leaves exactly half of the images from GO-12893 all of which 
by design used identical dither patterns, and exposure times that
were designed to take the primary target to the same count level.
Since the F555W and F775W images were designed to reach the same
exposure level, we do not separate detection completeness by filter.
The background level is nearly identical by design in using post-flash
across all cases.  The number of cosmic rays will of course vary 
frame-to-frame, but the longest total integration time in any filter
was 673 seconds summed over 5 exposures, and cosmic rays have
been well eliminated in the drizzling process and so are not a major factor.
Nor do we take into consideration differences in stellar crowding 
image-to-image, while at the \eke pixel scale of 4{\farcs}0 many of
these are quite crowded fields, at the \ehste WFC3 scale of 0{\farcs}04
per pixel these are all sparse.  While minor deviations in completeness
level surely exist image-to-image, this is likely below the level of
fidelity to which any reasonable effort could quantify this.

We have chosen to determine completeness on a relatively fine 
grid of offset positions from the primary targets: 0{\farcs}10,
0{\farcs}12, 0{\farcs}16, 0{\farcs}2, 0{\farcs}3, 0{\farcs}4,
0{\farcs}6, . . . 1{\farcs}8, and positions at and larger than 
2{\farcs}0.  Simulated images are generated at a 
series of delta magnitudes using the PSF appropriate (Section 3.1)
to each image.  The simulated images are always placed at random
positions with respect to sub-pixel centering.  The positions of
real stars, other than the primary target, in the real-data difference images were 
ignored in placing the simulated stars.  Different approaches to
accumulating a sufficient number of simulations were adopted for
different offset positions.  Outside of 2{\farcs}0 simulated 
stars in each of the 22 images were placed on a regular grid
(set at random sub-pixel position for each realization)
every 25 pixels in x and y per image, thus resulting in over 50,000 cases.
For the smallest offsets the central 41$\times$41 pixels in each
difference image were rastered into standard sized images, thus 
replicated over 900 times per image, and one simulated star was
added at the correct radial offset for the trial, but at a random
position angle.  For intermediate offsets a few realizations were
placed on the offset circle at non-interfering separation in angle.
A sufficient number of simulations were used to result in 
Poisson counting statistics errors of $\lesssim$1\% for the smallest
and largest offset positions up to $\sim$5\% for a few intermediate
offset separations.

Our simulated images use difference images from which the 
target has been subtracted as the basis.  For small offsets
and magnitude differences a concern existed that the presence
of a nearby companion might influence the primary target fit and subtraction,
thus invalidating use of difference images as the basis.
We therefore also performed a large number of simulations
based on injections to the original images, and found no
difference with respect to conclusions using the difference images.

After setting up the simulated images at any offset and delta magnitude 
combination the images were processed with {\tt DAOFIND}, and
the same cuts on output parameters as used on the real images were applied.
A target returned from {\tt DAOFIND} was counted as real if it falls
within 0{\farcs}07 of the known position of the simulation input.

\begin{figure}
\begin{center}
\includegraphics[width=80mm]{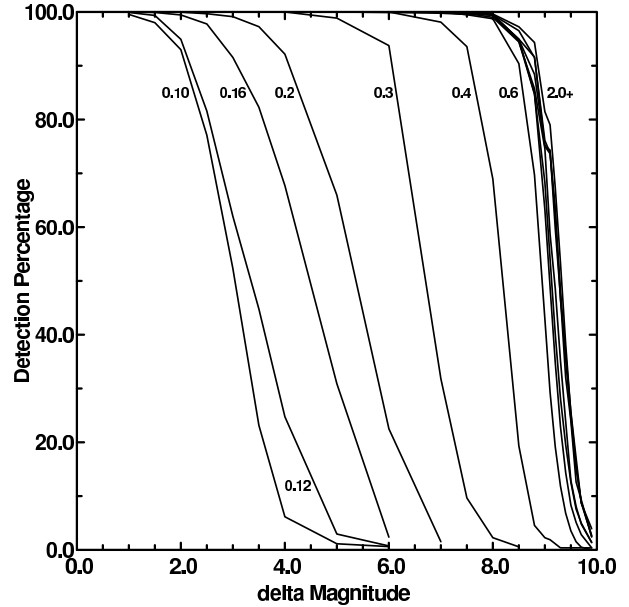}
\end{center}
\caption{Detection completeness in percent plotted against delta magnitude
for extensive simulated companions added to real images as discussed in
the text.  The curves are at different radial offsets as labelled in \arcsec.
Beyond 0{\farcs}8 curves are spaced every 0{\farcs}2, but beyond this there
is little change in sensitivity.}
\label{fig:complete}
\end{figure}

The resulting completeness levels are shown in Figure~\ref{fig:complete}.
At an offset of 0{\farcs}10, 90\% of stars with a delta magnitude of 2.0
are detected, while the 50\% recovery level occurs at about $\delta$-mag of 3.
At and outside of 2\arcsec, 90\% completeness is at $\delta$-mag 8.9
and 50\% at 9.4 per filter.
Clearly beyond an offset of 0{\farcs}8 the completeness levels change
very little, with most of the sensitivity change for offsets $\lesssim$0{\farcs}5.

We believe that at $\gtrsim$0{\farcs}2 these completeness levels have
been reached with essentially no chance of false positives coming in.
That view is based on having performed careful visual inspections
of all such detections in both F555W and F775W images and assuring
that all detections are qualitatively secure.
At yet lower offsets, we expect few if any false positives, but cannot
argue that none have occurred in the real data given non-Gaussian 
behavior of the difference images in the PSF cores, or at least PSF
variations that are not adequately captured by our modelling.

\begin{figure}
\begin{center}
\includegraphics[width=75mm]{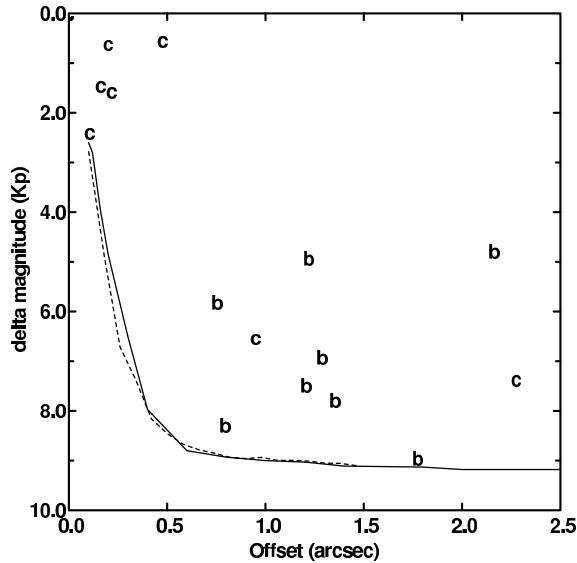}
\end{center}
\caption{All companion detections over the 23 targets within 2{\farcs}5 radial
offset are plotted with a Kp $\delta$-magnitude relative to the target.
Detections determined to be physically associated with the target are 
shown with `{\bf c}', and likely background, chance associations are 
shown with `{\bf b}' -- see Section 4.2 for discussion.
The curve is developed as the 70\% recovery level for each radial offset
as plotted in Figure~\ref{fig:complete}.
The image FWHM is $\sim$0{\farcs}078.  The dashed line showing mean
noise in annuli near the bright targets, represented as a magnitude 
loss is discussed in Section 3.4.}
\label{fig:sensdet}
\end{figure}

Figure~\ref{fig:sensdet} shows a sensitivity curve developed as the 
70\% completeness level (following this use in the \citet{lill14}
comparative study) for all offsets as plotted in Figure~\ref{fig:complete},
as well as all detections within 2{\farcs}5 offsets shown as
$\delta$-magnitude deviations from the target Kp.
The dashed line in Figure~\ref{fig:sensdet} corresponds to the boosted
magnitude near bright stars (as discussed in Section 3.2) based
on the noise in annuli spanning 0{\farcs}1 to 2\arcsec~ in steps of 
0{\farcs}08 near the bright targets in the PSF-subtracted images.
The dashed line is 2.5$\times$log($noise/reference$), where $reference$
is the mean noise level of 9.0 electrons per drizzled pixel
at $>$1{\farcs}5.

\subsection{Comparison with Other High Resolution Imaging}

Although the high-resolution imaging observations available under CFOP in support of 
vetting and interpreting \eke exoplanet candidates is expanding
rapidly, these \ehste observations fill a useful niche.
These observations are of very interesting targets in general,
and are unique in covering both a large 40$\times$40\arcsec~ domain,
and providing resolution better than 0{\farcs}08 in well-calibrated optical bandpasses
that can be robustly mapped to the \eke bandpass.

A recent paper by \citet{lill14} compared all primary published
results for high-resolution imaging of \eke targets.
Included were the authors own lucky-imaging results picking off
the best 10\% of very short, diffraction limited exposures obtained
with the 2.2 m telescope at Calar Alto Observatory, near-IR Adaptive
Optics results of \citet{adam12}, \citet{adam13}, and \citet{dressing14} using the 6.5 m
Multiple Mirror Telescope on Mt. Hopkins, speckle imaging \citep{howe11} with the 
3.5 m WIYN telescope on Kitt Peak and \citep{horc12} with the 
8 m Gemini North telescope on Mauna Kea, as well as the extensive
robotic AO observations (optical) obtained by \citet{law13} with
the 1.5 m Palomar telescope on Mt. Palomar.
We will also comment on Keck 10 m telescope on Mauna Kea near-IR 
adaptive optics imaging, which is extensively used in the \eke
literature, although lacking an over-arching summary paper.

An ideal set of high-resolution imaging observations would include
all KOIs, include a large enough area (40\arcsec~ would be good) to
cover all potential companions of interest for analyzing KOIs
while at the same time building up observational statistics on 
background object density across the \eke field, reaching limiting depths
of 9--10 magnitudes beyond the target brightness in order to detect
(with comfortable margin) all possible sources of false positives (shallower okay for deeper
transit cases), use at least two well-calibrated bandpasses in the
optical allowing robust transformation to Kp and interpretation
with published stellar isochrones, and have near diffraction limited
resolution on large telescopes to probe as near the target as possible.
All existing high-resolution surveys to analyze the \eke targets
fail in at least one of the attributes of an ideal program.
The primary weakness of our \ehste program is that it provides
data for only 23 ($<$1\% of total sample) KOIs, but is probably 
better than any other program on the balance of other desired attributes.
Therefore these \ehste results may provide useful insights 
through comparison of over-lapping observations from other programs.

A valuable observational element available through CFOP for 
nearly all KOIs are UKIRT near-IR images and resulting source tables.
However, those were based on rather shallow, natural seeing imaging.
In Figure~\ref{fig:img3049} we have highlighted two detections
at offsets and position angles from the KOI of about 14\arcsec~
and 203 degrees, and 19{\farcs}7 and 149 degrees.  In the UKIRT-based
source tables both of those were flagged as probably galaxies,
and had Kp magnitudes projected from $J$ of 21.5 and 21.0 respectively.
On the much deeper, much higher resolution \ehste images both 
objects are stellar in appearance and have more accurately
determined Kp magnitudes of 24.5 and 21.5 respectively.  

We have only one target in common with the 6.5 m AO \citep{dressing14},
the unusually faint for MMT AO, KOI-886 at Kp = 15.85.
The MMT AO provides a primary working angle of 0{\farcs}3 -- 10\arcsec~
to a contrast usually better than 5 magnitudes in $K_s$.
Within the 10\arcsec~ radius where \citet{dressing14} report no 
detections we see four.  
Three of these have $\delta$Kp $\gtrsim$8, while a fourth has 
an offset angle from the target of 3{\farcs}7, $\delta$Kp = 3.86 and
is fairly blue.  None of these should have been reported from the 
MMT AO given their listed sensitivity limits for this faint star.

We have six targets in common with Robo-AO \citep{law13}:  KOIs
854, 886, 1085, 1422, 1598, and 2045.  Within the typical offset 
working angle of $\sim$0{\farcs}2 -- 2{\farcs}5 and limiting depth of about 6 magnitudes for
Robo-AO no detections were made for these six targets.  For the same set
we have three detections within the same offset distance.
The KOI-1422 $\delta$-Kp 1.56 companion at 0{\farcs}22 was not found
by Robo-AO that deeper \ehste imaging reveals.
This is a very interesting target \citep{dres13}
hosting multiple planets in or near the habitable zone, that will
be strongly modified in interpretation \citep{cartier14} through the 
recognition that this is a stellar binary.  This is near
the advertised working domain for Robo-AO,
but in the low-performance domain for this faint KOI detection was not expected.
Also not found was the 2{\farcs}3 offset $\delta$-Kp of 7.36 
physical companion to KOI-1598 -- this is fainter than the 
claimed Robo-AO limit.  Similarly the 1{\farcs}3 $\delta$-Kp of 6.9
detection near KOI-2045 is not detected by Robo-AO.
Although strong in covering many (eventually all is the goal)
KOIs, the Robo-AO program by design does not succeed in detecting a large
number of relevant companions.

We have no overlap with the lucky imaging \citep{lill14}, but
note that the sensitivity curve of Figure~\ref{fig:sensdet} for \ehste
is much better than lucky imaging, especially at offsets $<$1\arcsec,
see their Figure 11.

\citet{lill14} discuss the limitations of the extensive
3.5 m speckle imaging \citep{howe11} which has neither a field
of view large enough, nor a detection depth deep enough to provide
much of an effect in delimiting possible target contamination.
We have KOIs 812 and 3259 in common with the more capable
Gemini 8 m speckle imaging discussed in \citet{horc12}.
For neither of these targets does the speckle imaging result
in any detections.  For KOI-3259 the \ehste imaging turns up
a companion at 1{\farcs}36 just inside the speckle field of 
view, but at a $\delta$-Kp of 7.8 is some 3.8 magnitudes beyond
the 5-$\sigma$ limiting depth as quoted on CFOP for either of 
these Gemini speckle targets.  For the two targets 
in common with our \ehste program, the Gemini speckle imaging
appears relatively uninteresting from the perspective of field
of view and limiting depth.  However, the speckle imaging 
provides the best resolution limit of any other available options
in common use.

J. Lillo-Box provided evaluation of the Blended source 
confidence (BSC) parameter used in \citet{lill14} for
our \ehste observations.  The BSC parameter measures how
effectively high resolution imaging observations eliminate
the phase space of background objects bright enough to 
be false positives.  \citet{gill11} had introduced a 
similar parameter and estimated that \ehste observations
would eliminate 95--98\% of possible background sources.
The \citet{lill14} BSC result came out quite similar
with typical rejections above 99\%, and averaging well
above 95\%.  In contrast the other high resolution
imaging observation programs discussed in \citet{lill14}
succeed in eliminating only a small fraction of possible
background objects bright enough to matter.

Finally we note that several of our targets have been observed
with the Keck NIRC2-AO system of the 10 m Mauna Kea telescope, as
was used for all of the objects discussed in \citet{marc14}.
The Keck AO images generally have a FWHM some 30\% better than
the \ehste value, and reach comparable limiting depth in many uses.
However, the field of view, and limitation to near-IR bandpasses
introduces limitations relative to the ideal.  Our most challenging
target, KOI-2626 (see also Cartier et al., 2014), with a triple at the
0{\farcs}2 level, was first detected in Keck AO imaging, then in 
short order also observed with Gemini 8 m speckle and our \ehste
imaging.  In all three of these observation sets the triple is 
cleanly resolved.  Arguably the \ehste imaging is most valuable in
providing well-calibrated optical photometry supporting determination
of light fraction of each component in the \eke bandpass, as well as
matching up with published stellar isochrones.

\section{Testing for Physical Association}

Spatially close companions near and below the arcsec offset level
are common in our high-resolution imaging.  Since knowing whether
a given companion is a chance superposition or a star likely 
bound in physical association with the target influences interpretation
of planetary candidates, we seek a means of determining this nature.
Canonical approaches would be to check for common proper motions, or
similar radial velocities -- neither is available in this study.
We know only the angular separation, and photometry in 
two \ehste bandpasses with WFC3, F555W and F775W.  Since the companions
are fainter than the targets, and many of our targets are already 
K and M dwarfs, if physically bound the companions will often
be M dwarfs.  We will use matching of closely separated stars 
to isochrones, arguing that if the multiple stars can be 
represented by a single isochrone, then they are likely a bound
system.  We also adopt a Bayesian argument providing odds ratios in
individual cases for the associations being a physical companion
versus chance alignment.

\subsection{Use of Isochrones -- M dwarfs and Optical Bandpasses}

If two stars can be placed on a common stellar evolution isochrone,
then this is solid evidence (in an Occam's razor sense) that the
two objects are at nearly the same distance,
likely share similar ages and metallicities, and hence are likely to
be physically bound.  The Dartmouth stellar evolution isochrones
\citep{dott08, feid11} conveniently include magnitudes in our
WFC3 bandpasses, and have been extensively used for the interpretation
\citep{dres13} of \eke M-dwarf KOIs.  Modeling of M-dwarfs is 
challenging: especially problematic is obtaining the correct 
spectral energy distribution at short wavelengths where the 
complexity of molecular opacities couples with a part of the 
spectrum containing only a small fraction of the stellar flux.

\begin{figure}
\begin{center}
\includegraphics[width=75mm]{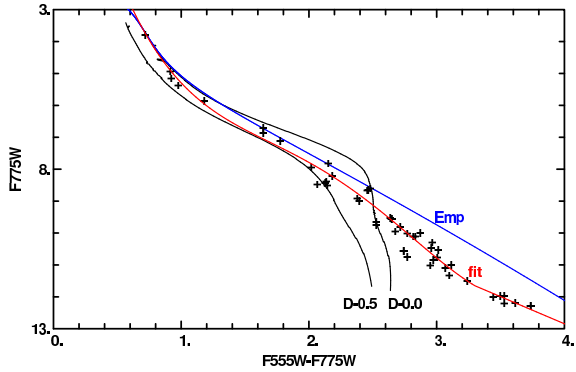}
\end{center}
\caption{Photometry of local stars from RECONS project
after minor transform of $V$ and $I_c$ to F555W and F775W.
Superposed are Dartmouth \citep{dott08, feid11} isochrones for [Fe/H] = -0.5
and 0.0, labelled D-0.5 and D-0.0 respectively.  The (blue) curve labelled
``Emp" was formed from synthetic photometry of \citet{pick98} spectra.
The (red) curve labelled ``fit" is a polynomial fit to the local stars.}
\label{fig:isochrone}
\end{figure}

Figure~\ref{fig:isochrone} illustrates the problems arising for 
interpretation of early-to-mid M dwarfs, roughly those beyond 
F555W-F775W of 2.0, using standard isochrones.
In this figure about 50 very nearby stars with $V$, $I_c$ 
photometry, and large, accurately known parallaxes from the 
RECONS project \citep{henr99, henr06, cant13, jao14} are shown
following application of a minor transformation to F555W and F775W
based on synthetic photometry using the \citet{pick98} spectra
and the IRAF/STSDAS {\tt SYNPHOT} package.
The Dartmouth
isochrones at [Fe/H] = 0.0 and -0.5 perform reasonably well in
matching the spectral energy distribution of earlier type stars,
but fail spectacularly in the optical with representing early 
to mid-M dwarfs (M3 is at F555W - F775W $\sim$2.4).
See also similar comparisons and more extensive
discussion in \citet{boya12}.
Recognizing this disconnect for the M dwarfs
that will likely be the most common companions in these analyses,
we used the \citet{pick98} set of standardized, composite spectra
to develop synthetic photometry using the {\tt SYNPHOT} package
under IRAF/STSDAS.  This empirical approach to setting an isochrone
results in a curve that at least parallels the observed color-magnitude
distribution in Figure~\ref{fig:isochrone}, but by mid-M is
offset by over a magnitude.

For our purposes we only care about whether the stars at small
spatial offsets are at near equal distances, and thus likely
to be physically associated.  We are not attempting to derive 
accurate stellar parameters.  Thus we are free to adopt a purely
empirical fit within the F775W, F555W-F775W CMD to represent 
a mean solar neighborhood isochrone.  
The adopted fit shown in Figure~\ref{fig:isochrone}
is a fourth order polynomial over F555W-F775W =
0.5 to 3.3 with a linear extension beyond this.  Although there
is a significant scatter of individual stars around this empirical
fit, likely due to generally unknown star-to-star metallicities,
this represents the empirical CMD much better than more direct
options such as the Dartmouth isochrones, or our attempt to
obtain an empirical isochrone from published composite spectra.

Figure~\ref{fig:isochrone} should raise a strong note of caution 
in using the Dartmouth isochrones to interpret M dwarfs if relying
on optical bandpass photometry.
The reddest stars plotted in Figure~\ref{fig:isochrone} 
extend only to M6.

\subsection{Odds Ratios for Companion versus Chance Alignment}

We follow the approach used by \citet{torr11} and \citet{fres11} to establish
the probability that planetary candidates are real exoplanets,
instead of false positives.  In their case an odds ratio based 
upon the established frequency of exoplanets, and the frequency
of potential false positives was established.  In our case this
will be very similar -- a ratio based on the known physically-associated
companion rate, compared to our own empirically developed
rate of false associations:

\begin{equation}
{\rm Odds~Ratio} = \frac{N_{expected~bound}}{N_{random~alignments}}
\end{equation}

The recent synthesis of an extensive
literature on the commonality of binary and higher order associations
by \citet{duch13} is adopted to provide the prior expectation on 
frequency of bound systems and thus $N_{expected~bound}$ for each target.
\citet{duch13} present results that
FGK dwarfs have a mean companion frequency of 0.62, while M dwarfs
have a lower fraction of 0.33 companions per star.  For the G dwarfs
companions follow a log-normal distribution with mean semi-major
axis of 45 AU and a $\sigma$ on log(Period) in days of 2.3, 
and 5.3 AU and $\sigma$log(Period) of 1.3 for M dwarfs.  

Our high-resolution imaging is sensitive primarily to binaries
at relatively large physical separations, the high period tail of
the log-normal distributions.  To evaluate the statistically expected
number of companions for each of our targets we proceeded as follows:
(1) For each target, estimate its distance using the observed F775W
magnitude, the reddening given in the KIC \citep{brow11}, and the 
empirical fit to absolute F775W versus color as in Figure~\ref{fig:isochrone}.
These distances are shown in Table 1.
(2) For each companion map the observed
spatial offset and estimated distance into AU ignoring projection effects.
(3) Integrate over the G and M distributions of companion frequency
separately from 0.5 to 2.0 times the offset distance in AU.
(4) If the spectral type of the target is G use the G star result,
if M the M star result, and if K the geometric mean of G and M.
We use the KIC $\teff$ from Table 1 and 5250 K as the G -- K boundary,
and 3850 K \citep{pick98} as the K -- M boundary.
The expected number of companions, $N_{expected~bound}$, within the search range is usually
0.02 -- 0.08, i.e., an order of magnitude smaller than
the integral number of companions expected on average over all possible offsets.

Assessing the $N_{random~alignments}$ is developed from our own
stellar detections and depends on three factors:  (1) The mean
stellar density (number per unit area) as a function of galactic latitude.
(2) The mean stellar density as a function of apparent brightness.
(3) How often unassociated stars reach various levels of consistency
with isochrone matching.
For the odds ratio of companion versus chance association we used
the full set of GO-12893 images to establish the frequency at which
two physically-unrelated stars match to a given $\chi^2$ in:

\begin{eqnarray}
\chi^2 & = & [(\Delta \rm{(F555W-F775W)}_A/\sigma_A)^2 \\ 
 & + & (\Delta \rm{(F555W-F775W)}_B/\sigma_B)^2 \nonumber \\ 
 & + & (\Delta \rm{F775W}_{B-A}/\sigma_{B-A})^2]/2 \nonumber
\end{eqnarray}
\noindent
where the color differences are F555W-F775W per star minus the same 
quantity for the fit to local stars as in Figure~\ref{fig:isochrone}
for the two stars tested, A and B, and the magnitude difference is 
the observed B-A value minus the corresponding value of the absolute
F775W from the fit.
For each of the 23 primary targets test matches were made to each 
of the non-primary targets in the other 22 images.
For each $\chi^2$ evaluation a minimum was evaluated
by testing against all colors in the underlying fit of Figure~\ref{fig:isochrone}.
In addition the observed colors were dereddened by all possible values 
(in steps of 0.01) from zero to twice the KIC value for $E(B-V)$ of the primary star.

\begin{figure}
\begin{center}
\includegraphics[width=65mm]{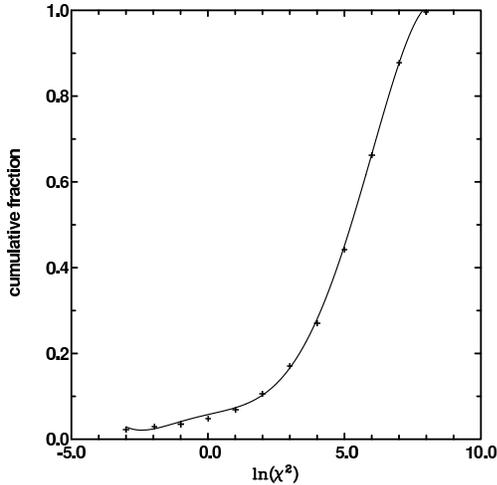}
\end{center}
\caption{Cumulative fraction of $\chi^2$ matches using Eq. (2) for randomly 
paired stars across frames that are not physically associated with each other.
Solid curve is a polynomial fit used in evaluating odds ratio for physical
association test.}
\label{fig:cumulativechi}
\end{figure}

From our set of 23 images, the smallest separation between two images
on the sky is about 0.75 degrees.  Our targets have a mean distance 
of $\sim$660 pc.  Thus the physical separation between stars in the
two images nearest each other would be of order 10 pc.  We thus 
assumed that no stars in any image are physically associated with 
stars in other images.  Using all detected stars in each of the 
23 images to evaluate Eq. (2) against all target stars in separate images
gave about 11,000 matches.  For the $\sigma$'s in Eq. (2) we adopted
formal estimates of precision from fitting a PSF to the data, with 
a floor of 0.015.  At 21st magnitude errors were $\sim$0.02, increasing
to typically 0.1 at 24th.  The $\chi^2$ comparison of two physically-unassociated
stars was usually large, with the most common values
of order 1000 indicative of extremely poor matches relative to the
empirical isochrone.  The cumulative fraction of matches as a function
of $\chi^2$ is shown in Figure~\ref{fig:cumulativechi}.  
Random pairings of stars do occasionally have good matches to the 
isochrone, e.g., 5\% have $\chi^2$ $\lesssim$1.0.  Since these stars
cannot be physically associated, it is clear that merely having a
close match with the isochrone does not suffice to argue for 
physical association.

\begin{figure}
\begin{center}
\includegraphics[width=65mm]{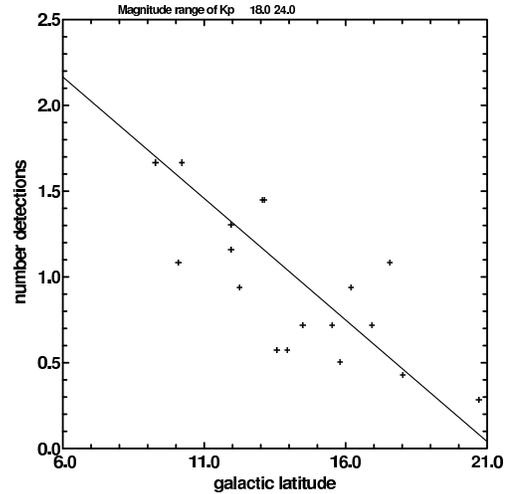}
\end{center}
\caption{Relative number of stars detected within Kp magnitude range of 18 -- 24 where
all images are complete to this depth plotted versus galactic latitude.
Number has been normalized by a mean of 13.7 stars per 1600 square arcsec image.
The straight line is a fit to the distribution to be used in adjusting
spatial density.}
\label{fig:gallat}
\end{figure}

\begin{figure}
\begin{center}
\includegraphics[width=65mm]{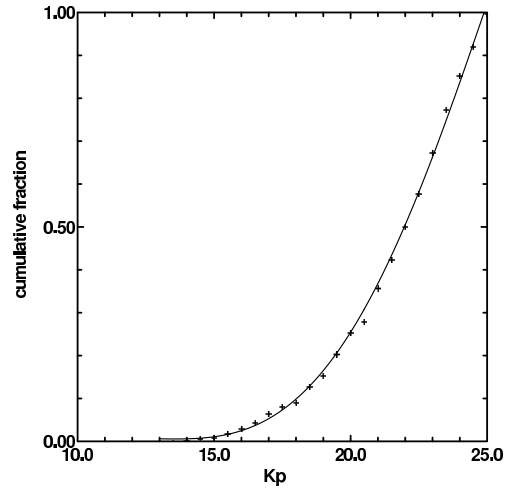}
\end{center}
\caption{Cumulative fraction of detected stars as a function of Kp over 
all images.}
\label{fig:fracbright}
\end{figure}

The test matching of $\sim$11,000 pairs covered an equivalent of
23$\times$22 times the 1600 square arcseconds of each image.
To evaluate the denominator for the odds ratio we adopted the same 
annulus area used for the prior expectation of companions discussed
above and multiplied this by the number density per unit solid angle of test
matches up to the $\chi^2$ level of the star in question.
We adjusted this value by two additional factors affecting the number
density of stars:  (1) a general dependence on galactic latitude
as shown in Figure~\ref{fig:gallat} after normalization to a mean
of unity, and (2) the dependence of number counts as a function of 
brightness Kp as shown in Figure~\ref{fig:fracbright} evaluated up
to the estimated brightness of the test star plus a 1$\sigma$ error.

This approach to evaluating an odds ratio for likelihood that a 
given companion is physically associated as opposed to a chance 
alignment provides a good indication of status, but for a few 
ambiguous cases additional inputs would be needed to establish the
true nature.  
The results of testing the 481 total detections over 23 images for
physical association is shown in Table~\ref{tab:physlist} for all
cases in which the odds ratio favors association, and for which the 
$\chi^2$ value is less than 10.  (Two cases had odds ratios of 
1.1 and 2.3 respectively for which the corresponding isochrone
$\chi^2$ were at 27 and 154.)

\begin{table}
\scriptsize
\begin{center}
\tablenum{3}
\caption{Companions with Likely Physical Association
\label{tab:physlist}}
\begin{tabular}{rrcrcccr}
\tableline\tableline
KIC & KOI & offset & PA & $\delta$-mag & color & $\chi^2$ & odds ratio \\ \hline
5358241 & 829 & 0.107 & 239.9 & 2.39 & 1.034 & 1.123 & 3703.9 \\ 
5358241 & 829 & 3.311 & 335.4 & 6.00 & 2.659 & 0.032 & 1.3 \\ 
6263593 & 3049 & 0.478 & 196.9 & 0.54 & 1.362 & 0.011 & 1923.7 \\ 
10004738 & 1598 & 2.275 & 65.9 & 7.36 & 2.933 & 0.038 & 6.3 \\ 
11497958 & 1422 & 0.217 & 217.3 & 1.56 & 2.478 & 0.008 & 4101.6 \\ 
11768142 & 2626 & 0.161 & 181.6 & 1.44 & 2.389 & 1.019 & 928.1 \\ 
11768142 & 2626 & 0.201 & 212.7 & 0.61 & 2.299 & 0.107 & 2832.9 \\ 
12256520 & 2264 & 0.949 & 103.8 & 6.53 & 2.838 & 0.038 & 62.5 \\ \hline 
\end{tabular}
\footnotesize
\tablecomments{KIC, and corresponding KOI numbers are given for all
stars having a positive odds ratio for physical association as discussed
in Section 4.2.  Offset is in arcseconds, PA in degrees E of N, $\delta$-mag in Kp,
color is F555W-F775W, with $\chi^2$ and odds ratio as in Section 4.2.}
\end{center}
\end{table}

The split in Figure~\ref{fig:sensdet} between dominance by 
companions within 0{\farcs}5 and chance alignments prevailing
outside of 0{\farcs}5 is striking.  
At the distances of our targets 0{\farcs}5 -- 2{\farcs}5 already
projects to the high-period tail of the log-normal, physically-associated distributions.
Given the values from \citet{duch13} we expect about twice as many physically
associated companions within 0{\farcs}5 as within 0{\farcs}5 -- 2{\farcs}5, i.e.,
consistent with Figure~\ref{fig:sensdet}.
The density of chance alignments should be uniform, and the area 
over 0{\farcs}5 -- 2{\farcs}5 is 25 times that at $<$0{\farcs}5.
In qualitative terms the split in Figure~\ref{fig:sensdet} is fully expected.

\subsection{Close Companions That Could Be Transit Host}

We now consider all detections which complicate consideration of
the true transit host based on our high resolution imaging.
For close companions that we have determined are likely in physical
association with the target the complication would be not knowing
the true host star of the planet candidates.
For chance alignments the nearby star remains as a false positive 
possibility.  In tabulating these cases of interest we consider
two factors:  (1) Whether the offset distance of the companion star
is within the 3-$\sigma$ centroiding error based on analysis of 
difference images in- and out-of-transit \citep{brys13}.
(2) Whether the companion star is bright enough such that even 
with 90\% eclipse depths it could produce the observed
transit depth when diluted by the target.

We take the 3-$\sigma$ centroiding error from the Data Validation
reports for Quarters 1-16 from CFOP.  Two entries from Table 3
are not included here, since for both the 3{\farcs}3 offset star
for KIC-5358241, and the companion to KIC-10004738 the \ek-based
centroiding eliminates these from contention.  On the other hand
we find that newly detected close companions to KIC-6263593 = KOI-3049
that are not physically associated with the target cannot be 
eliminated as possible false positive sources for the apparent transits.

\begin{figure}
\begin{center}
\includegraphics[width=85mm]{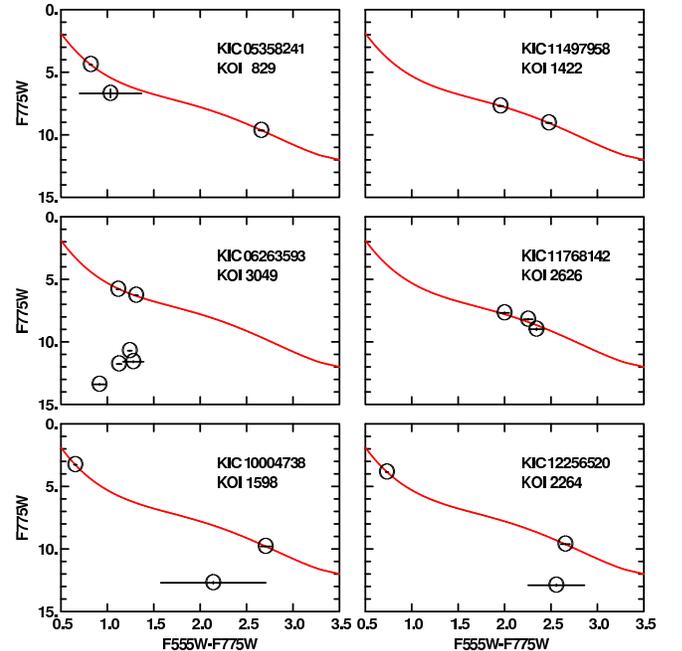}
\end{center}
\caption{CMDs illustrating the relative positions 
for all targets listed in Table 3, and all companions within 4\arcsec.
Target and companion are adjusted in color by dereddening of up 
to twice the KIC $E(B-V)$ to yield best fit.  The distinction
between likely physical associations and chance alignments is 
generally obvious.  The target is the
brightest star, and by design falls on the selected isochrone.
A circle is superposed on the location of each star.}
\label{fig:fig3049}
\end{figure}

The primary and companions for the 6 targets in Table 3 are shown in relation to 
each other in the CMD Figure~\ref{fig:fig3049}.
KOI-3049/KIC-6263593 provides a good illustration --
the closest companion in physical offset and $\delta$-mag has an odds
ratio of nearly 2000 favoring physical association.
The other three companions in Table 4 have odds ratios favoring
chance alignment and $\chi^2$ isochrone match values (Eq. 2) of 154 to 816.

\begin{table}
\scriptsize
\begin{center}
\tablenum{4}
\caption{Target in Which Host is Uncertain from High-Resolution Imaging
\label{tab:confounding}}
\begin{tabular}{rrcccrc}
\tableline\tableline
KIC & KOI & 3-$\sigma$ error & offset & $\delta$Kp & Depth if host & bound? \\ \hline 
5358241 & 829 & 0.57 & 0.11 & 2.39 & 4235 & yes \\ 
6263593 & 3049 & 1.39 & 0.48 & 0.54 & 1428 & yes \\ 
6263593 & 3049 & 1.39 & 0.76 & 5.82 & 115460 & no \\ 
6263593 & 3049 & 1.39 & 1.21 & 7.48 & 530684 & no \\ 
6263593 & 3049 & 1.39 & 1.22 & 4.92 & 50704 & no \\ 
11497958 & 1422 & 1.03 & 0.22 & 1.56 & 4098 & yes \\ 
11768142 & 2626 & 2.07 & 0.16 & 1.44 & 3899 & yes \\ 
11768142 & 2626 & 2.07 & 0.20 & 0.61 & 2253 & yes \\ 
12256520 & 2264 & 2.31 & 0.95 & 6.53 & 137027 & yes \\ \hline 
\end{tabular}
\footnotesize
\tablecomments{KIC, and corresponding KOI numbers are given for all
stars leading to ambiguity as to the host star.  The 3-$\sigma$ error is the
extreme offset still allowed from \eke centroid analysis, offset is in
arcseconds, the ``Depth if host" is the intrinsic transit or eclipse
depth (for shallowest transit of multi-planet systems)
in ppm if the star is the host, and the final column indicates
whether the companion is physically associated with the primary target star.}
\end{center}
\end{table}

Over the 23 KOIs considered in this paper we find that 5 have
physically-associated companions that could be the true transit host,
and one of these has additional chance alignments that could be the
source for a false positive.  The results are summarized in
Table~\ref{tab:confounding}.  Refined analyses, e.g., application
of BLENDER \citep{torr11}, might well eliminate some of these 
possibilities.  

The `Depth if host' in Table 4 follows from the observed depth
in an unrecognized blend and the correction for dilution.
\cite{daem09} provide the by now often used correction for dilution
in the case of the host being the brighter star.  With the assumed
host being the fainter star the equation takes the same form, with
a change of sign in the exponent:

\begin{equation}
{\rm Depth~if~host} = {\rm Depth}_{\rm obs} (1 + 10^{\delta {\rm Kp}/2.5})
\end{equation}

\noindent
For several cases the inferred depth if the companion is the host
exceeds 10\% making a planet interpretation unlikely, 
although this must be balanced by considering the host star
radius as well which is beyond the scope of this paper.

Of the stars listed in Table~\ref{tab:physlist} or Table~\ref{tab:confounding}
one deserves a special note of caution.  The companion to KIC-5358241 at an
offset of 0{\farcs}107 and a $\delta$Kp of 2.39 is in a region of detection
parameter space where reliability is difficult to assess.  Until this 
detection is confirmed by a second high-resolution imaging experiment, the 
reality of this star remains open to question.  All of the other detections
listed in these tables are considered unambiguously real with parameters as listed
within nominal uncertainties.

\subsection{Statistics on Physical Association}

As shown in Table~\ref{tab:physlist} we have detected six companions 
to five targets that have odds ratios above 50, which we take as 
confirmation of physical association.  Two additional cases have
positive indications of being bound to the target, but are not certain.
We wish to tabulate here the total number of detections as a function
of target spectral type and compare this with the {\em a priori} 
number of expected companions \citep{duch13}.  For the two cases with
modest odds ratios of 1.3 and 6.3 we assign fractional detections
as 1 - 1/(odds ratio + 1).
We also only count the KIC-5358241 source offset by 0{\farcs}107
as 0.5 of a detection from conservatism
as to the basic reality of its existence.

Over the 21 targets considered, ignoring the two sources
(KIC-6149553 and KIC-11306996) that in the interim had been shown to
be false positives,
we have 7, 8, and 6 stars at G, K, and M spectral types respectively.
To evaluate expected number of companions in our data we sum over the
log-normal distributions \citep{duch13} corresponding to search offsets
of 0{\farcs}12 to 22{\farcs}57.  This yields an expected total number
of companions of: 1.52, 0.67, and 0.31 for the G, K, and M stars in our
sample.  We detect 1.93, 2.0, and 3.0 across the three spectral types.
Overall we detect 6.93 compared to an expected number of 2.50, for 
an over-detection ratio of $\times$2.8.  The fraction of times that
Poisson fluctuations with an expectation of 2.5 exceed 7 is 1.4\%
suggesting that the general over-abundance of companions is significant
at $>$98\%.  For the M-dwarfs in isolation the over-abundance is $\times$9.7,
and the nominal significance is 99.6\%, or 3$\sigma$.

It is clear that (a) this experiment yields only small number statistics
with 6 -- 8 targets per spectral type bin, and (b) we find more physically-associated
companions than would be expected.  The over-abundance of 
physically-associated companions is a monotonically increasing factor
over the G (near nominal), to K, and M (order of magnitude excess)
spectral types.  These targets were selected on the basis of
being strongly suspected of hosting one or more small
and cool exoplanets.  A direct interpretation is that widely-spaced stellar 
binaries, and higher order systems are over-represented among 
planetary hosts.  Such an interpretation from this limited sample would be consistent with
stellar multiples being more favorable sites for exoplanet formation,
than are single stars.
Another factor is that with knowledge of binarity these planets are
actually larger than reported in the exoplanet archive.  There has
thus been a selection effect of including larger planets in the 
sample due to unrecognized binarity initially.

\citet{krau14}, however, find from an interferometric survey of 600
\eke planet candidate hosts sensitive to 5 -- 50 AU binaries
(our sensitivity is primarily to binaries near and mostly
beyond 50 AU, and therefore disjoint to their sample)
that stellar binarity has a strong influence {\em suppressing}
the planet frequency by a factor of about four.

G. W. Marcy (2014, private communication) suggests an alternative view that multiple
stars within the \eke aperture simply increases the probability 
of transits existing, e.g., doubled for a binary
since either of the stars might host planets at random orientations.
In this approach one might expect binaries to be over-represented
in the KOI list by a factor of two if the individual components
have the same intrinsic planet hosting frequency as single stars.
Since our over abundance at M-dwarfs is nearly a factor of 10
a discrepancy persists that simply interpreted suggests 
wide binaries favor the formation of small planets.

\section{Summary}

The stable, high-resolution imaging provided by WFC3 UVIS on \ehste
has allowed a careful survey for stellar companions to KOIs selected
for having small and cool exoplanet candidates.  In most cases no
close-in companions were found that could be false positives for 
transits through blending of background eclipsing binaries.  
Our sensitivity to such companions reached 90\% completeness
for delta-magnitudes of 4 at 0{\farcs}2 separation, 8 at 0{\farcs}4,
and $>$9 magnitudes at $\gtrsim$0{\farcs}8, with the latter usually
exceeding the general brightness ratio that could physically be 
a false positive for these transits.  However, in several cases 
we find close-in companions that are physically associated with 
the target star.  These smaller, associated stars would not likely
be a concern regarding potential false positives for exoplanets,
but do call into question which of the multiple stars hosts the
planet candidates with strong implications for the resulting
exoplanet properties as presented by \citet{cartier14}.

We also find, especially for our sample of M dwarf stars, that 
stellar multiplicity is over-represented in association with our
sample having been selected as likely exoplanet hosts.  Resolving whether 
the implied proclivity of wide multiple stars, as can be found in
high-resolution imaging surveys, to form exoplanets is merely a
fluke of small number statistics awaits larger studies.

\acknowledgements 

Funding for \ek, the tenth Discovery mission, was provided by NASA's Science Mission Directorate.
The many people contributing to the development of the {\em Kepler Mission}
are gratefully acknowledged.
The individuals who joined as co-investigators on this \ehste
proposal in early 2012 are recognized.
Andrea Dupree,
Francois Fressin,
Matthew Holman,
Jack Lissauer, 
Geoff Marcy,
and Jason Rowe
contributed directly to development of the proposal, provided 
extensive background discussion, or assisted in target selection.
We thank Jorge Lillo-Box for providing note of his detection 
metric for our \ehste imaging.
R.L.G. and K.M.S.C. have been partially supported through grant {\em HST}-GO-12893.01-A
from STScI.
P.K. thanks support from NASA NNX11AD21G and NSF AST-0909188.
The Center for Exoplanets and Habitable Worlds is supported by the 
Pennsylvania State University, the Eberly College of Science, and 
the Pennsylvania Space Grant Consortium.
CFOP is funded by NASA through the NASA Exoplanet Science Institute.
Data presented in this paper were obtained from the 
Mikulski Archive for Space Telescopes.

{\it Facilities:} \facility{{\em HST} (WFC3)}, \facility{{\em Kepler}}.

\end{document}